\documentclass[aps,prl,twocolumn,letterpaper,superscriptaddress,showpacs]{revtex4}
\usepackage{graphicx}
\usepackage{CJK}
\usepackage{color}
\usepackage[normalem]{ulem}

\begin{document}
\begin{CJK*}{UTF8}{bsmi}
\title{Effects of disordered Ru substitution in BaFe$_2$As$_2$: Realization of superdiffusion mechanism in real materials}
\author{Limin Wang}
\affiliation{Condensed Matter Physics and Materials Science Department,
Brookhaven National Laboratory, Upton, New York 11973, USA}
\author{Tom Berlijn}
\affiliation{Condensed Matter Physics and Materials Science Department,
Brookhaven National Laboratory, Upton, New York 11973, USA}
\author{Yan Wang}
\affiliation{Department of Physics, University of Florida, Gainesville, FL 32611, USA}
\author{Chia-Hui Lin}
\affiliation{Condensed Matter Physics and Materials Science Department,
Brookhaven National Laboratory, Upton, New York 11973, USA}
\affiliation{Physics Department, State University of New York, Stony Brook,
New York 11790, USA}
\author{P. J. Hirschfeld}
\affiliation{Department of Physics, University of Florida, Gainesville, FL 32611, USA}
\author{Wei Ku}
\affiliation{Condensed Matter Physics and Materials Science Department,
Brookhaven National Laboratory, Upton, New York 11973, USA}
\affiliation{Physics Department, State University of New York, Stony Brook,
New York 11790, USA}

\date{\today}

\begin{abstract}
An unexpected insensitivity of the Fermi surface to impurity scattering is found in Ru substituted BaFe$_2$As$_2$ from first-principles theory, offering a natural explanation of the unusual resilience of transport and superconductivity to a high level of disordered substitution in this material.
This robustness is shown to originate from a coherent interference of correlated on-site and inter-site impurity scattering, similar in spirit to the microscopic mechanism of superdiffusion in one dimension.
Our result also demonstrates a strong substitution dependence of the Fermi surface and carrier concentration, and provides a natural resolution to current discrepancies in recent photoelectron spectroscopy.
These effects offer a natural explanation of the diminishing long-range magnetic, orbital and superconducting order with high substitution.
\end{abstract}


\maketitle
\end{CJK*}

Chemical substitution is the most widely employed technique to induce high-temperature superconductivity~\cite{iron1, iron2, iron3, RPP,Canfield}.
The most obvious effects of substitution are carrier doping and chemical pressure.
On the other hand, such substitution should also introduce disordered impurity scattering, an effect mostly unexplored in the studies of electronic structure of real materials due to its complexity.
Recently, owing to an advances in first-principles theoretical methods~\cite{unfolding, disorder}, various novel and surprising physical effects of disorder were found in the new Fe-based superconductors containing transition metal dopants~\cite{Co, Co2, TB_Co} and Fe vacancies~\cite{vacancy}.
Naturally, one would wonder whether there are also unexpected disorder effects in
isovalent substitution.
An ideal candidate to investigate is the Ru substituted Fe-based superconductor Ba(Fe$_{1-x}$Ru$_x)_{2}$As$_{2}$(Ru122)~\cite{ARPES1Ru,ARPES2Ru,ARPES3Ru,ARPES4Ru,transport_Ru,phasediagram,Singh,Nakamura,VCA_Ru}.
Unlike other existing cases with isovalent substitution at the anion sites, Ru atoms replaces the most essential Fe atoms.
Given that Ru is a 4d element, it must be quite different chemically from 3d Fe and therefore must introduce a strong impurity scattering potential.

One obvious physical puzzle regarding Ru122 is the resilience of its superconductivity and transport against large concentration of disordered impurities.
Indeed, samples are found to retain their superconductivity even with 40\% Ru substitution of Fe~\cite{transport_Ru,ARPES2Ru}.
Given the current proposal that the superconducting order parameter is most likely of $s_\pm$ symmetry~\cite{RPP}, it is extremely puzzling how a sign-changing order parameter can survive such a large concentration of strong impurities.
Similarly, given that Ru is more distinct from Fe than Co is, it is quite unexpected that Ru substituted samples exhibit a residual resistivity comparable to 8\% Co substituted system~\cite{Canfield}at a much higher 35\% substitution level.

Another important current issue is the Ru substitution dependence of the electronic structure.
Qualitatively similar to the Co substituted systems, Ru substitution at high enough level systematically suppresses the magnetic, orbital order, and then superconductivity.
For Co substituted systems, this behavior can be understood from the weakening of the nesting between the electron and hole pockets, since the former grow in size while the latter shrink, responding to the additional doped carriers~\cite{TB_Co}.
For Ru substitution, however, the size of the electron and hole pockets should remain balanced (and approximately nested), since Ru is supposed to be isovalent to Fe.
It is thus not as obvious what substitution-dependent features in the electronic structure suppress  the long-range order in this case.

Currently, this issue is quite controversial in the field, due to seemingly contradictory experimental observations.
Some angular resolved photoemission spectroscopy (ARPES) experiments~\cite{ARPES2Ru} reported a nearly substitution-independent Fermi surface, and suggested that superconductivity emerges from dilution of the magnetism, in contrast to other doped 122 systems.
{ Other} ARPES measurements, on the other hand, reported a large increase of the number of carriers~\cite{ARPES1Ru} with increasing Ru concentration and a crossover from two-dimensional to three-dimensional structure of some of the hole-like Fermi surfaces~\cite{ARPES3Ru}.
Similar contradictions also emerge in current first-principles computations.
Some concluded that Ru substitution induces no doping~\cite{Singh} and no changes in the low energy band structure~\cite{Nakamura}, while others~\cite{VCA_Ru} found large changes.
It is thus timely to investigate the substitution-dependence of the electronic structure of Ru122 with a proper account of the disorder effects to resolve the current debates, and if possible to offer an explanation for the suppression of long-range magnetic, orbital and superconducting order upon increased substitution levels.

In this Letter, we address these important issues by studying the electronic structure of Ba(Fe$_{1-x}$Ru$_x$)$_2$As$_2$ over the full range of substitution ($0\leq x\leq 100$) taking into account the realistic disorder effects in first-principles calculation.
This is made possible via the recently developed Wannier function based effective Hamiltonian method for disordered systems ~\cite{disorder}.
Surprisingly, while large scattering is found on the entire Fe band complex, for all substitution levels, the states near the chemical potential remain very sharp and coherent.
This unexpected insensitivity of the quasiparticles to impurity scattering is here traced back to a coherent interference between on-site and off-site impurity effects.
In spirit, this is very similar to the microscopic mechanism that gives rise to superdiffusion in 1D theory~\cite{superdiffusion}, and can thus be considered as the first realization of this exotic phenomenon in real materials.

This insensitivity to impurity scattering provides a natural explanation of the amazing resilience of transport and superconductivity in the presence of a high level of Fe site substitution in this specific material.
In addition, our results reproduce the measured spectral functions and resolve the current controversy in the interpretations of the Ru substitution dependence.
We find a systematic reduction of carrier density upon Ru substitution accompanied by an enhanced 3D character of the Fermi surface.
These effects lead naturally to suppression of long-range magnetic and superconducting order at high substitution level.
These findings highlight the general need to incorporate disorder effects in most chemically substituted systems.

\begin{figure*}
\includegraphics[scale=1]{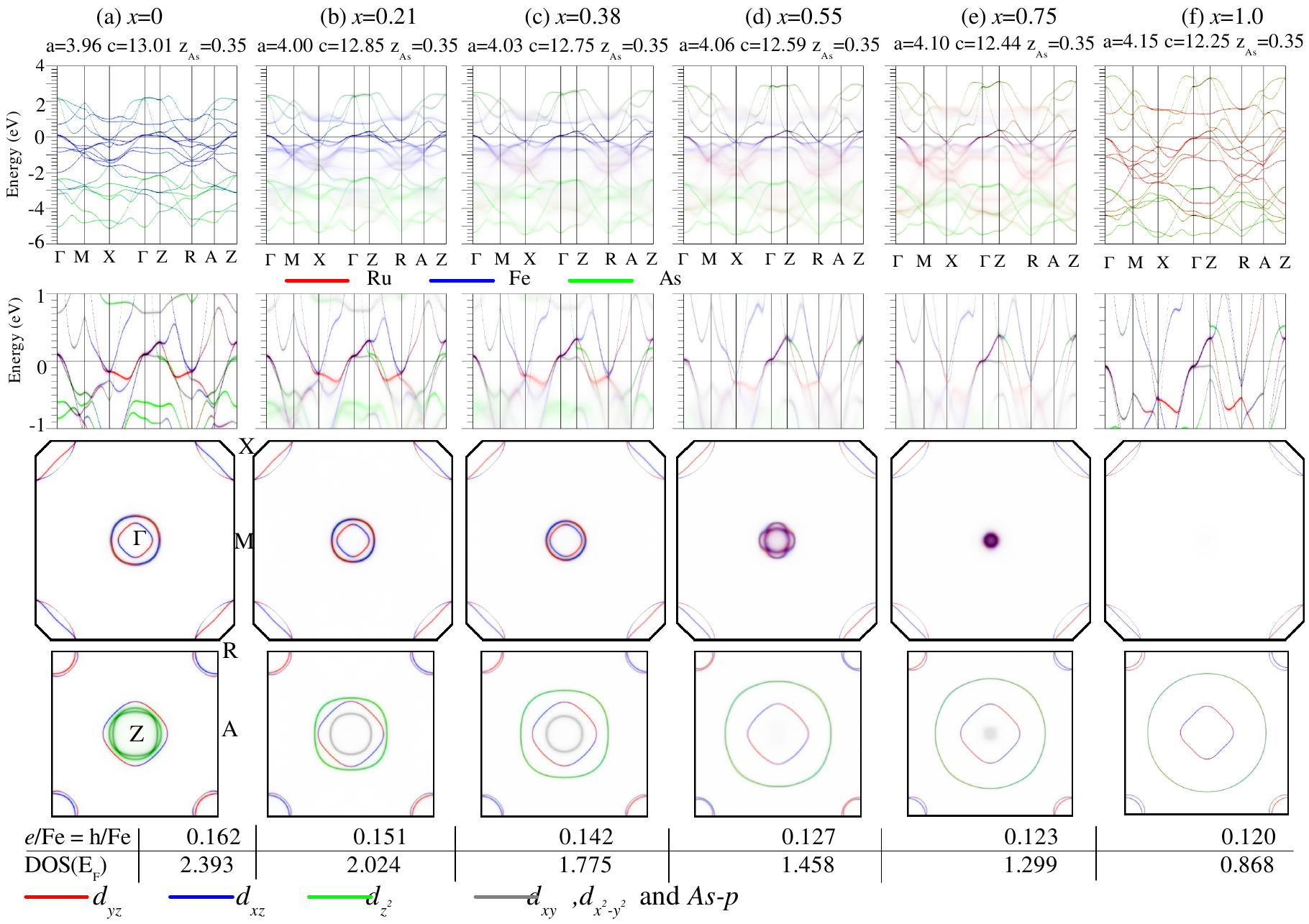}
\caption{(color online) The lattice constants ({\AA})~\cite{transport_Ru,0,1}, band structure (the first panel), low energy band structure(the second panel), the Fermi surface around $\Gamma$ point with $k_z=0$ (the third panel) and around $Z$ point with $k_z=\pi$ (the last panel) for Ba(Fe$_{1-x}$Ru$_x$)$_2$As$_2$ with different Ru concentration $x=$ 0 (a), 0.21 (b). 0.38 (c), 0.55 (d), 0.75 (e) and 1.0 (f). Orbital characters of Fe, Ru, and As are distinguished by color given in the legend.  Resulting carrier density per Fe, and density of states at the Fermi level $N(E_F)$ per (eV Fe) are given below the plots.}
\label{fig:fig1}
\end{figure*}

We first evaluate the configuration-averaged spectral function $\langle A_{n}(k,\omega)\rangle$ of Wannier orbital $n$, frequency $\omega$, and crystal momentum $k$,
within the approach of Ref. \onlinecite{disorder}, and
with the help of the band structure unfolding method~\cite{unfolding}.
The resulting $\langle A_{n}(k,\omega)\rangle$ is found converged within the resolution of the plots after averaging over 10 large randomly-shaped supercells containing 400 atoms on average with random Ru substitutions.
The influence of Ru substitution on the Hamiltonian is extracted from the results of several  density function theory (DFT) calculations: the undoped BaFe$_2$As$_2$ or BaRu$_2$As$_2$, as well as the impurity supercells of Ba$_2$Fe$_3$RuAs$_4$ or Ba$_2$FeRu$_3$As$_{4}$.
The low energy Hilbert space is taken within [-6,4] eV consisting of Wannier orbitals of Fe-$d$, Ru-$d$ and As-$p$ characters.
Careful benchmarks of the quality of our effective Hamiltonian are conducted against full DFT calculations using stoichiometric ordered ``impurity lattice" test cases~\cite{sup}.
The entire study is conducted using the 2-Fe Brillouin zone required by the symmetry~\cite{1FeVs2Fe}, and unfolded to the 1-Fe zone~\cite{unfolding,1FeVs2Fe} when direct comparison with ARPES spectra is needed.


Figure~\ref{fig:fig1} summarizes our resulting spectral functions over the entire range of substitution for Ba(Fe$_{1-x}$Ru$_{x}$)$_2$As$_2$ ($x=0.0, 0.21, 0.38, 0.55, 0.75$ and $1.0$).
A clear broadening of quasiparticle spectral lines in both momentum and frequency is observed over the entire Fe $d$-band complex, reflecting the finite mean free path and lifetime of carriers in these states due to disorder.
The broadening is naturally the strongest for the $x=0.55$ case, since it is the most disordered one shown.
(As $x$ approaches 0 or 1, the system starts to become cleaner, approaching the pure end compounds.)
Such a strong impurity scattering is expected, since the 4d level of Ru is about 0.8 eV lower in energy than the 3d level of Fe, and the different spatial extent of the orbitals  also implies large changes in their hopping strength.

Strikingly, the bands near the Fermi level (shown in the second panel of Fig. 1) remain very sharp, as if carriers occupying states near the Fermi level don't scatter from the impurity.
Correspondingly, the configurational averaged spectral function [Fig. 2(a)] exhibits a very sharp peak at the Fermi level with a well-defined dispersion.
(In comparison, around $\sim -1$ eV the spectral function shows a much broader peak, reflecting strong impurity scattering effects.)
Therefore, the puzzling resilience of transport and superconductivity against high substitution level in Ru122 is now understandable simply from the lack of net impurity scattering for states near the Fermi level.

Fig.~\ref{fig:fig2}(b)-(d) give further insights into the microscopic origin of this unusual insensitivity to disorder near the Fermi level.
Here we compare spectral functions under the influence of different partial components of the impurity potential in the Wannier function basis.
With only the diagonal impurity potential (on-site disorder), Fig.~\ref{fig:fig2}(c) shows a strong smearing of the entire $d$-band complex, including states near the Fermi level.
On the other hand, with only the off-diagonal impurity potential (off-site disorder), Fig.~\ref{fig:fig2}(d) shows a well structured smearing of the band structure: the farther away from the center of the $d$-bands, the more the states are affected by the impurity.
In particular, near the center of the band, the effects of the off-site impurity potential diminishes.
This is understandable since the off-site terms of the Hamiltonian are responsible for the band dispersion, and their effects is maximum at the band edges.
One would expect a similar structure for the effectiveness of off-site impurity potential, particularly positive/negative above/below the center of the band.
(The Ru orbital is bigger than Fe, so the impurity potential tends to enhance the band width.)
A similar conclusion can be reached
by imagining the band structure  interpolated between pure BaFe$_2$As$_2$ and pure BaRu$_2$As$_2$.

Upon combining both on-site and off-site disorder potential, Fig.~\ref{fig:fig2}(b) shows that the energy range with weak impurity scattering moves up, close to the Fermi level.
This can now be understood as a consequence of the cancelation of (negative) on-site and (positive) off-site impurity potential above the center of the band.
Indeed, Fig.~\ref{fig:fig2}(e) gives an example of the net coupling involving 8th/7th and 9th/10th bands near the $\Gamma$ point resulting from a single impurity for all three cases.
One sees that near the Fermi level, the impurity scattering between 9th and 10th bands has large on-site contribution (similar to that between 8th and 7th bands), but it is almost entirely canceled by the equally large off-site scattering.
It is important to note that such coherent interference can only take place because the positions of on-site and off-site impurity potential are always correlated: both are associated with the location of Ru substitution.
Therefore, moving the impurity position only adds the same overall phase to both of the terms but does not affect their interference.

This effect is in spirit very similar to the microscopic mechanism that gives rise to the unusual superdiffusion in 1D models~\cite{superdiffusion}.
Due to the poor connection between sites, (non-interacting) 1D systems are known to be extremely sensitive to disorder.
Any tiny on-site disorder leads immediately  to a total localization of the system.
Surprisingly, it can be shown~\cite{superdiffusion} that the system can still overcome the localization to give a ``superdiffusive'' metallic transport if certain special correlations exists in the position of the impurities to ensure a coherent interference of their scattering.
What we find here in Ru122 is probably a more general (and realistic) case, in which the interference takes place at a single impurity, but between on-site and off-site contributions.
 From this point of view, Ru122 can thus be considered the first realization of such an exotic superdiffusion mechanism in real materials.
This realization manifests itself with a significant consequence that superconductivity survives to a high substitution level, due to the insensitivity of states near the chemical potential to the impurity scattering.
In general, similar interference should occur in some other materials with disordered impurities as well, since realistic impurity induced on-site and off-site effects are always correlated in position.
This might also explain
why metallic transport can be found in some thin nanowires~\cite{nanowire1, nanowire2} and quasi-1D materials~\cite{Q1D1, Q1D2} , despite the unavoidable occurrence of disorder.

\begin{figure}[tbp]
\includegraphics[scale=0.5]{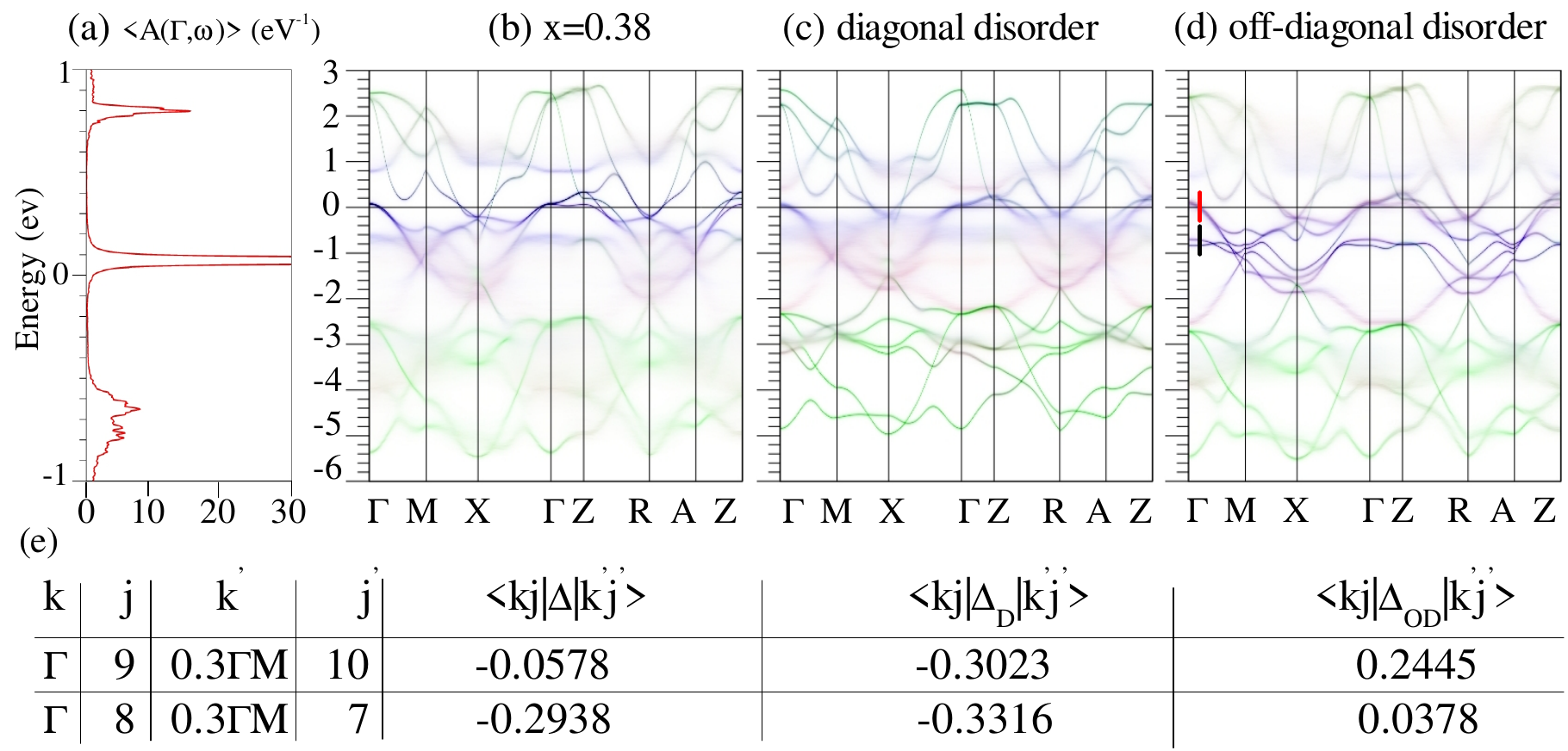}
\caption{(color online) Spectral function (a) and band structure for system with two types of disorder (b), with only the diagonal disorder (c), and with only the off-diagonal disorder(d) for disordered Ba(Fe$_{0.62}$Ru$_{0.38}$)$_2$As$_2$. (e) single-impurity induced coupling between the Bloch orbitals near and far from the chemical potential, indicated by the black and red segments in (d).}
\label{fig:fig2}
\end{figure}

Next, considering Ru substitution dependence of the electronic structure, Fig.~1 show several clear trends.
First, the hole pockets in the center of the Brillouin zone becomes more three-dimensional, shrinking significantly in size in the $k_z=0$ plane and eventually disappearing, while surviving in the $k_z=\pi$ plane.
Specifically, the pocket with large $z^2$ character shows the strongest 3D dispersion, growing in size in the $k_z=\pi$ plane.
On the other hand, the electron pockets near the corner of the zone shows only a slight substitution dependence.
This large change is quite consistent with the recent ARPES observation~\cite{ARPES3Ru}, but \textit{seems} to contradict  the earlier observation of substitution independent Fermi surface~\cite{ARPES2Ru}.

This discrepancy can, in fact, be easily resolved by accounting for the matrix element of the incident photon in the experiment~\cite{ARPES_orbital}.
The photon polarization used in Ref.~\cite{ARPES2Ru} is along the $y$ direction, which couples mostly to the $yz$ orbitals.
Figure~3 shows the $yz$ character of our theoretical unfolded band structure~\cite{unfolding}, showing little change upon Ru substitution at the $k_z=\pi$ plane, in excellent agreement with the experiment.

\begin{figure}[tbp]
\includegraphics[scale=0.5]{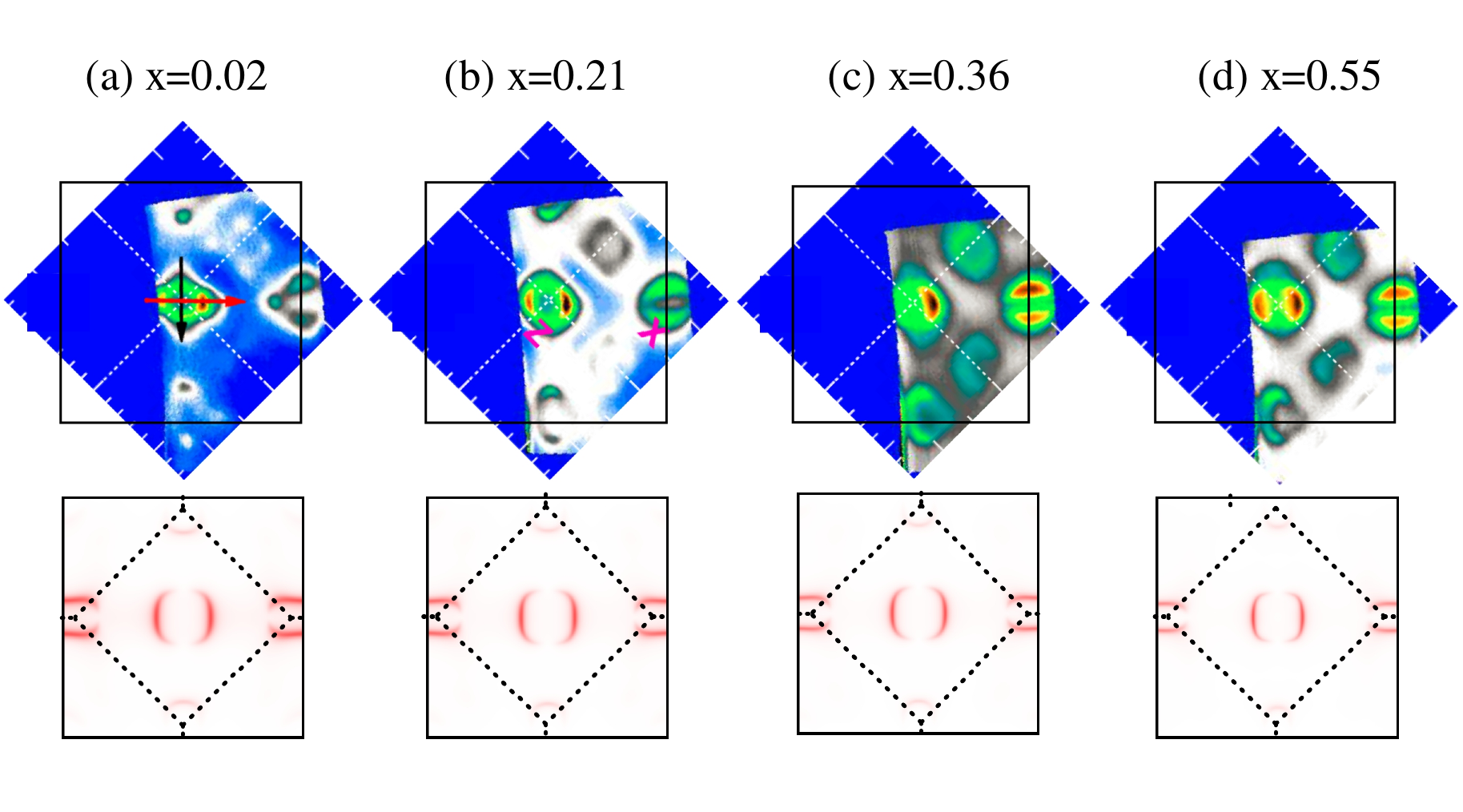}
\caption{(color online) Comparison of the Fermi surface at $k_z=\pi$ plane between the ARPES results~\cite{ARPES2Ru} (top panel) and our results (bottom panel) in 1-Fe zone for different Ru concentration $x=$ 0.02 (a), 0.21 (b), 0.36 (c) and 0.55 (d). The dashed lines marks the true 2-Fe Brillouin zone boundaries.}
\label{fig:fig3}
\end{figure}

The second physical effect of Ru substitution revealed in Fig.~1 is an physical reduction of carrier density, a real doping effect, and correspondingly a reduction of the density of states (DOS) at the chemical potential.
In typical systems with only one
sign of carriers, carrier doping is typically achieved via substitution of different valence.
Here, Ru substitution, while typically considered isovalent, actually reduces the carrier concentration significantly, from 0.16/Fe at $x=0$ to 0.12/Fe at $x=1$, about a 25\% reduction (c.f. bottom of Fig.~1.)
This takes places through reducing the 3D volume of both the electron and the hole pockets (while keeping charge neutrality), as if the system is being drained of both electron and hole simultaneously.
Microscopically, this originates from a larger splitting between the conduction and valence bands, due to the enhanced hopping via Ru substituted sites.
Such  chemical ``pressure'' effects should survive the mass renormalization observed in real materials.

Our result, which is perfectly consistent with the recent ARPES measurement~\cite{ARPES3Ru}, suggests a  more complex picture.
Indeed, we found that the green $z^2$ pocket increases in size at $kz=\pi$ upon Ru substitution, as observed experimentally~\cite{ARPES3Ru}.
However, the total volume of the hole pockets (or equivalently that of the electron pockets) actually reduces, as reflected in the corresponding carrier density shown at the bottom of Fig.~1.

The above substitution dependence of the electronic structure corresponds nicely to the overall features in the phase diagram of this material, from the standard weak coupling perspective.
Both the enhanced 3D nature and the reduced DOS tend to suppress the long-range magneic/orbital order, and eventually the superconductivity with sufficient high substitution level (the ``overdoped'' regime.)
Indeed, the enhanced 3D band structure weakens the nesting condition for magnetic/orbital correlation, and presumably the glue for superconductivity in the ``overdoped'' regime.
Similarly, the smaller DOS reduces the Fermi surface instability in both the magnetic, orbital, and pairing channels.
In essence, the case of Ru substitution offers a unique test case with strong substitution dependence of the electronic structure but without tipping the balance between electron and hole pockets.
This unique feature of Ru122 should serve as a good qualitative benchmark for all proposed theories of superconductivity.


In summary, a remarkable insensitivity of the states near the chemical potential to disordered impurity scattering is found in Ba(Fe$_{1-x}$Ru$_x)_{2}$As$_{2}$, despite the strong impurity potential of Ru.
This offers a natural explanation of the unusual resilience of transport and superconductivity against a high impurity level.
This exotic behavior originates from a coherent interference of on-site and off-site disorder potential, and thus can be considered the first realization of super-diffusion mechanism in real materials.
Our result, while being very consistent with existing ARPES measurement, resolves a few current discrepancies in the field, and leads to new physical conclusions.
Specifically, Ru substitution is found to physically reduce the carrier density, instead of being ``iso-valent''.
Our findings offer a natural explanation of the diminishing long-range magnetic, orbital and superconducting order at high substitution level.

The authors thank Vladimir Dobrosavljevic for pointing out the similarity of our mechanism to that of the superdiffusion.
Work funded by the U S Department of Energy, Office of Basic Energy Sciences DE-AC02-98CH10886, DE-FG02-05ER46236 and DOE-CMCSN.

\begin{widetext}

\section{[Supplementary Information]}

\subsection{Details of the density functional theory calculations}\label{sec:secdft}

\begin{figure*}[htp]
\includegraphics[width=0.7\columnwidth]{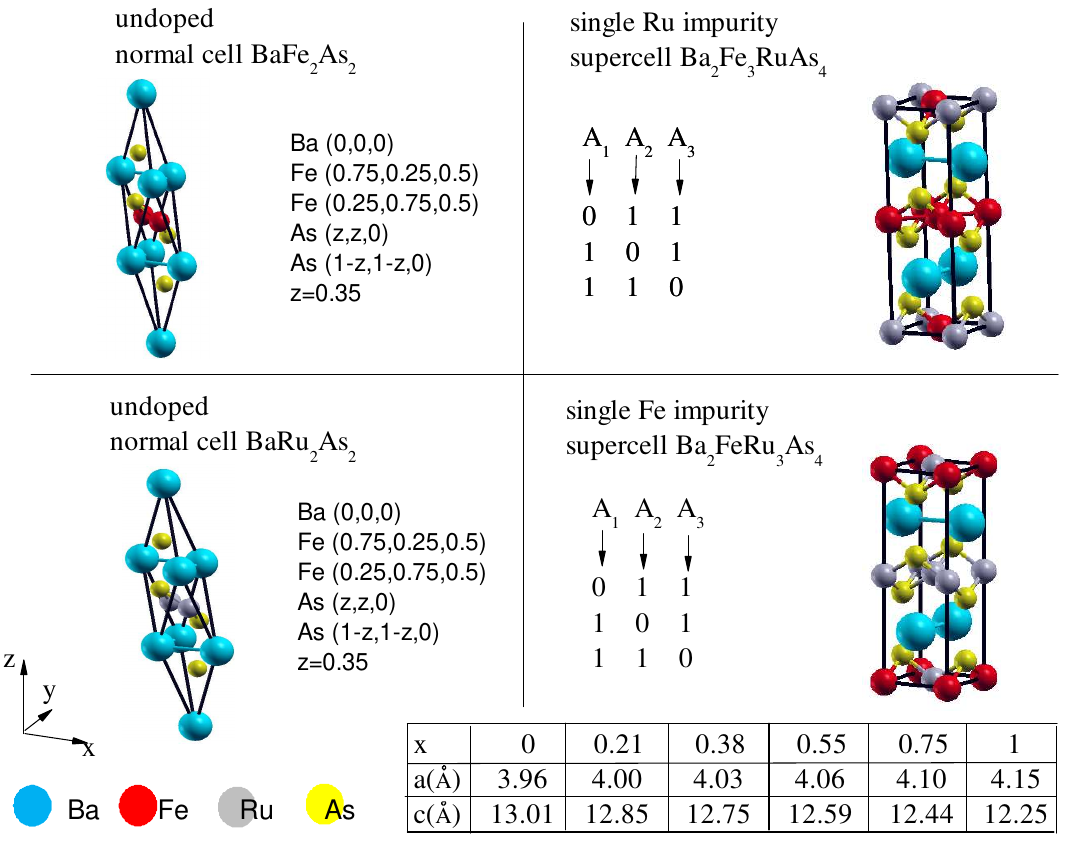}
\caption{\label{fig:fig1}
(Left panel) Undoped normal cell BaFe$_2$As$_2$/BaRu$_2$As$_2$ , and (right panel) single Ru/Fe impurity supercell Ba$_2$Fe$_3$RuAs$_4$/Ba$_2$FeRu$_3$As$_4$, and (right bottom) lattice constants used for the different doping levels.}
\end{figure*}
We investigate the electronic structure of Ba(Fe$_{1-x}$Ru$_x$)$_2$As$_2$ for the six doping levels listed in the table at the right bottom of figure \ref{fig:fig1}.
For the end points $x=0$ and $x=1$, ordinary Density Functional Theory (DFT) is used, while for the doping levels in between the effective Hamiltonian method~\cite{naxco2} with the help of the band structure unfolding method~\cite{unfolding} is employed to be able to properly treat the influence of disorder.
As an input for the effective Hamiltonian method~\cite{naxco2} we perform two DFT calculations for each of the four intermediate doping levels.
For $x=0.21$ and $x=0.38$ we use BaFe$_2$As$_2$ as the undoped system and treat Ru as an impurity, while for $x=0.55$ and $x=0.75$ we use BaRu$_2$As$_2$ as the undoped system and treat Fe as an impurity.
The lattice structures of the undoped normal cells, BaFe$_2$As$_2$/BaRu$_2$As$_2$ , are depicted in the left panel of figure \ref{fig:fig1}.
The primitive normal cell lattice vectors expressed in Cartesian coordinates are given by:
$ a_1=-\frac{1}{2}a\hat{x}+\frac{1}{2}a\hat{y}+\frac{1}{2}c\hat{z} \; ; \;  a_2=\frac{1}{2}a\hat{x}-\frac{1}{2}a\hat{y}+\frac{1}{2}c\hat{z} \; ; \; a_3=\frac{1}{2}a\hat{x}+\frac{1}{2}a\hat{y}-\frac{1}{2}c\hat{z}$.
The lattice parameters for the doping levels $x=0$, $x=0.38$, $x=1$ are taken from\cite{0,0.38,1}, respectively, and those for the doping levels $x=0.21$, $x=0.55$, $x=0.75$ are obtained from linear interpolation between $x=0$ and $x=1$.
To capture the influence of the Ru/Fe impurity a two times larger supercell is used to properly treat the non-local influence on the nearest As and Fe/Ru orbitals.
The corresponding Ba$_2$Fe$_{3}$RuAs$_{4}$/Ba$_2$FeRu$_{3}$As$_{4}$ supercells are depicted on the right panel of figure \ref{fig:fig1} and its primitive lattice vectors expressed in normal cell lattice vectors are given by:
$ A_1=a_2+a_3 \; ; \;  A_2=a_1+a_3  \; ; \; A_3=a_1+a_2 $.
We applied the WIEN2K\cite{blaha} implementation of the full potential linearized augmented plane wave method in the
local density approximation.
The k-point mesh was taken to be 11$\times$11$\times$11 for the undoped normal cell and 17$\times$17$\times$5 for the supercells respectively.
The basis set sizes were determined by RKmax=7.

\subsection{Benchmarks of the effective Hamiltonian against DFT}

To explore the accuracy and efficiency of the effective Hamiltonian method ~\cite{naxco2} for the case of Ba(Fe$_{1-x}$Ru$_x$)$_2$As$_2$, we present comparisons of spectral functions $A_n(k,\omega)$  calculated from the full DFT and the effective Hamiltonian (see figures \ref{fig:fig2}-\ref{fig:fig3}). The size of the deviations between the full DFT and the effective Hamiltonian should be compared with the size of the impurity induced changes. For this purpose the spectral function of the undoped BaFe$_2$Se$_2$/BaFe$_2$Ru$_2$ is also plotted as a reference for each benchmark. The Ba$_4$Fe$_4$Ru$_4$As$_8$ supercell in figures \ref{fig:fig2} and \ref{fig:fig3} is designed to test the linearity of the impurity influence (see formula (1) of Ref.~\cite{naxco2}) and the partitioning of the impurity influence from its super images (see section IV of the supplementary material of Ref.~\cite{naxco2}).
The basis set of Linear Augmented Plane Waves (LAPW's) used in the full DFT is $\sim30$ times larger then the basis set of Wannier functions used in the effective Hamiltonian method. Since the number of floating point operations of diagonalization depends cubically on the size of the matrix this implies an efficiency increase by a factor of $30^{3}\approx3\cdot10^{4}$. Furthermore the full DFT calculation involves multiple self consistent cycles (typically 15) whereas the effective Hamiltonian method (as currently implemented) requires only a single diagonalization, which increases the efficiency by another order of magnitude to $\sim 6\cdot10^{5}$.

\begin{figure}[htp]
\includegraphics[width=0.8\columnwidth]{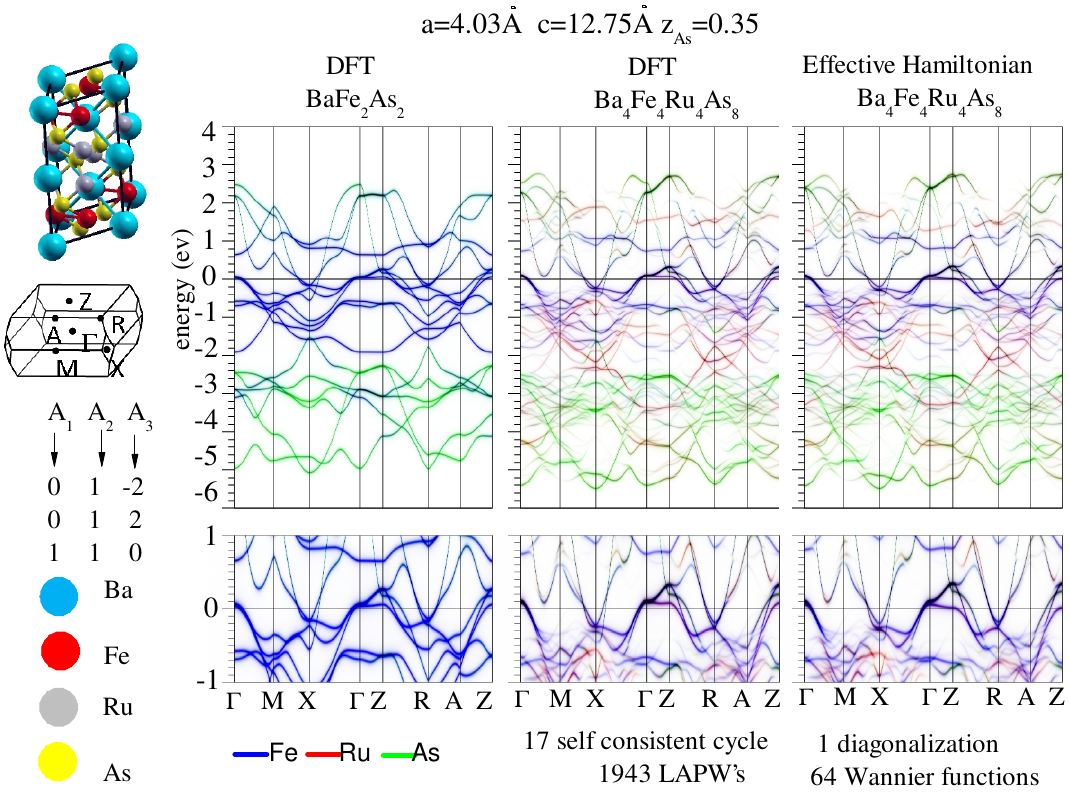}
\caption{\label{fig:fig2}
}
\end{figure}

\clearpage

\begin{figure}[htp]
\includegraphics[width=0.8\columnwidth]{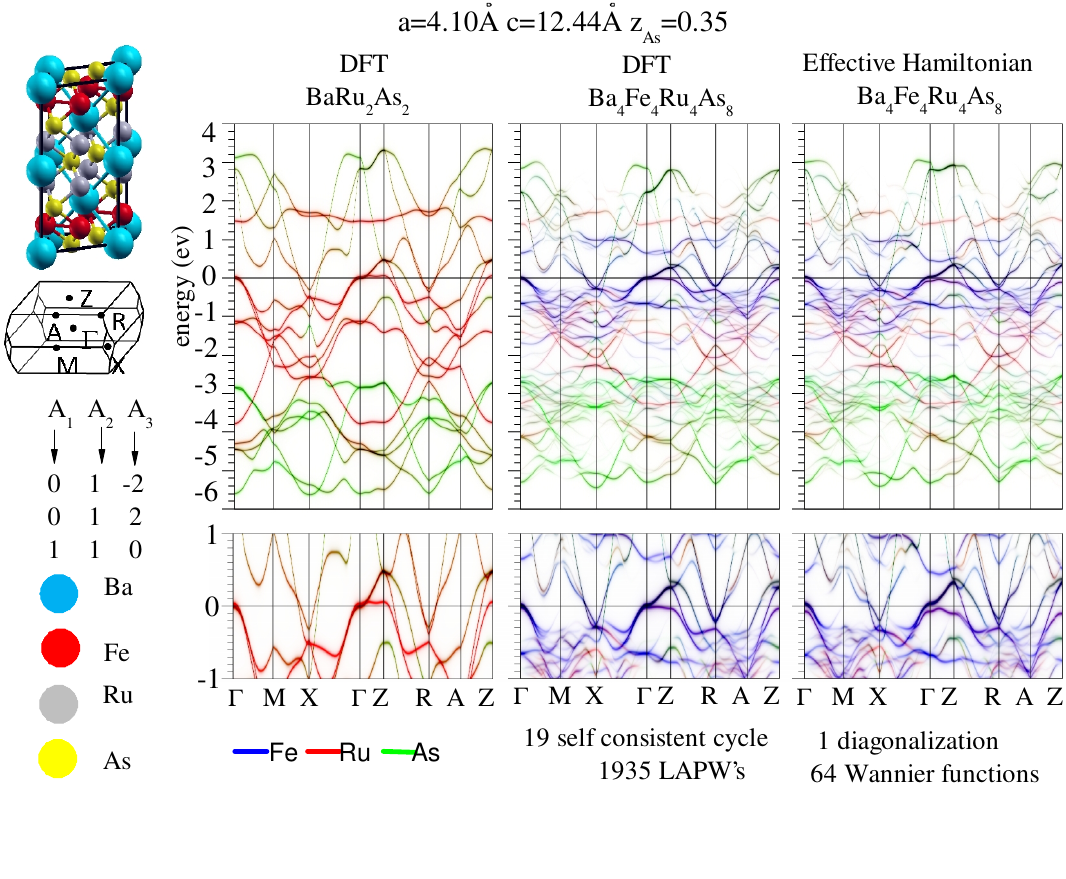}
\caption{\label{fig:fig3}
}
\end{figure}

\end{widetext}
\end{document}